\documentclass{ws-p8-50x6-00}

\begin{document}

\title{Lattice Calculation of Baryon Masses using the Clover
Fermion Action}

\author{D.G.~Richards}

\address{Jefferson Laboratory, 12000
Jefferson Avenue, Newport News, VA 23606, USA}
\author{M.~G\"{o}ckeler, P.E.L.~Rakow}

\address{Institut f\"ur Theoretische Physik, 
Universit\"at Regensburg, D-93040 Regensburg, Germany} 

\author{R.~Horsley, C.M.~Maynard}

\address{Department of Physics \& Astronomy, University of
  Edinburgh, Edinburgh EH9~3JZ, Scotland, UK}

\author{D.~Pleiter, G.~Schierholz}

\address{Deutsches Elektronen-Synchrotron DESY,
             John von Neumann Institute for Computing NIC/Deutsches
	     Elektronen-Synchrotron DESY,
                    D-15738 Zeuthen, Germany}

\maketitle 

\abstracts{We present a calculation of the lowest-lying
baryon masses in the quenched approximation to QCD.  The calculations
are performed using a non-perturbatively improved clover fermion
action, and a splitting is found between the masses of the nucleon and
its parity partner.  An analysis
of the mass of the first radial excitation of the nucleon finds a
value considerably larger than that of the parity partner of the
nucleon, and thus little evidence for the Roper resonance as a simple
three-quark state.}

\section{Introduction}
The calculation of the excited nucleon spectrum provides a theatre to
explore many of the central questions in hadronic physics, including
the applicability of the quark model, the r\^{o}le of excited glue,
and the existence of ``molecular'' states.  Recently, several
lattice calculations of the masses of lowest-lying nucleon states have
appeared, using a variety of fermion
actions.~\cite{lee98,lee00,sasaki99,adelaide02} In this talk, I describe a
calculation of the mass of the lowest-lying negative-parity state
using an ${\cal O}(a)$-improved clover fermion action. By using
a variety of volumes and lattice spacings, we are able to estimate
finite-volume and finite-lattice-spacing effects; further details of
this calculation are provided in earlier papers.~\cite{lat00,lhpc01}
For a subset of our lattices, we also determine the mass of the first
radial excitation of the nucleon.

\section{Calculational Details}
There are two interpolating operators that we will consider in our
measurement of the low-lying $J=1/2$ nucleon spectrum:
\begin{eqnarray}
N_1^{1/2+} & = & \epsilon_{ijk} (u_i^T C \gamma_5 d_j) \nonumber
u_k\label{eq:N1},\\ N_2^{1/2+} & = & \epsilon_{ijk} (u_i^T C d_j)
\gamma_5 u_k\label{eq:N2}.
\end{eqnarray}
These operators have an overlap with particles of both positive and
negative parity; on a lattice periodic or anti-periodic in time, the
best delineation that can be achieved is that of a forward-propagating
postive-parity state, and a backward-propagating negative-parity one.

The ``diquark'' piece of $N_1$ couples upper, or large,
spinor components whilst that of $N_2$ couples an upper and a lower
spinor component and hence vanishes in the non-relativistic limit.
Thus we expect $N_1$ to have a better overlap with the positive-parity
ground state than $N_2$.  The expectation is that $N_2$ couples
primarily to the lightest radial excitation of the nucleon, which
experimentally is the so-called Roper resonance $N^*(1440)$.

The calculation is performed in the quenched approximation to QCD,
using the the standard Wilson gluon action and the non-perturbatively
improved ``clover'' fermion action.  The quark propagators are
computed using both local and smeared sources.  Where possible, errors
on the fitted masses are computed using a bootstrap procedure, but
simple uncorrelated $\chi^2$ fits are employed in the chiral
extrapolations.

\section{Results}
The masses of the nucleon and its parity partner are obtained from
four-parameter fits to the two-point
correlators of $N_1$.
For the chiral extrapolation of the masses, we adopt the ansatz
\begin{equation}
(a m_X)^2 = (a M_X)^2 + b_2(a m_{\pi})^2
\end{equation}
where $X$ is either $N^{1/2+}$ or $N^{1/2-}$.  The leading
non-analytic term in the quenched approximation is linear in
$m_{\pi}$,~\cite{labrenz96} but results for $a M_X$ are insensitive to
this term, and indeed in the case of $N^{1/2+}$ we find a coefficient
whose central value differs in sign from that predicted.

\begin{figure}[t]
\begin{center}
\epsfxsize=220pt
\epsfbox{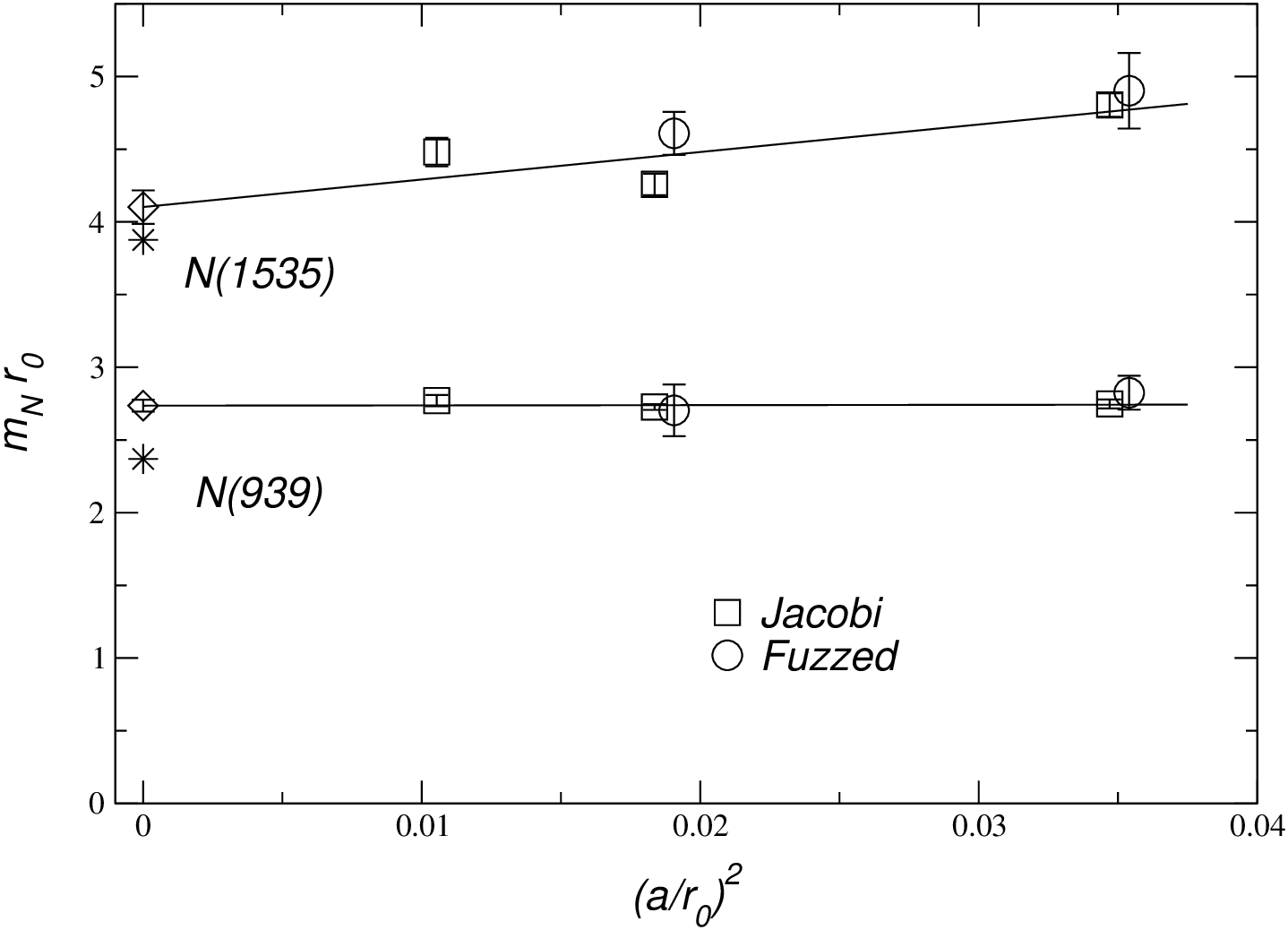}
\caption{Masses of nucleon and its parity partner in units of
$r_0^{-1}$ where $r_0 \sim 0.5~ \mbox{fm}$. The labels ``Jacobi'' and
  ``Fuzzed'' refer to two different nucleon smearing techniques used
  to improve the signal for the ground-state masses.}
\label{fig:extrap}
\end{center}
\end{figure}

In order to compare our data to experiment, we show in
Figure~\ref{fig:extrap} the masses of the nucleon and its parity
partner at each lattice spacing; we find good consistency between the
lattice calculation and the physical values, despite systematic
uncertainties due to the chiral extrapolation, finite-volume and the
use of the quenched approximation.

\begin{figure}[t]
\begin{center}
\epsfxsize=140pt
\epsfbox{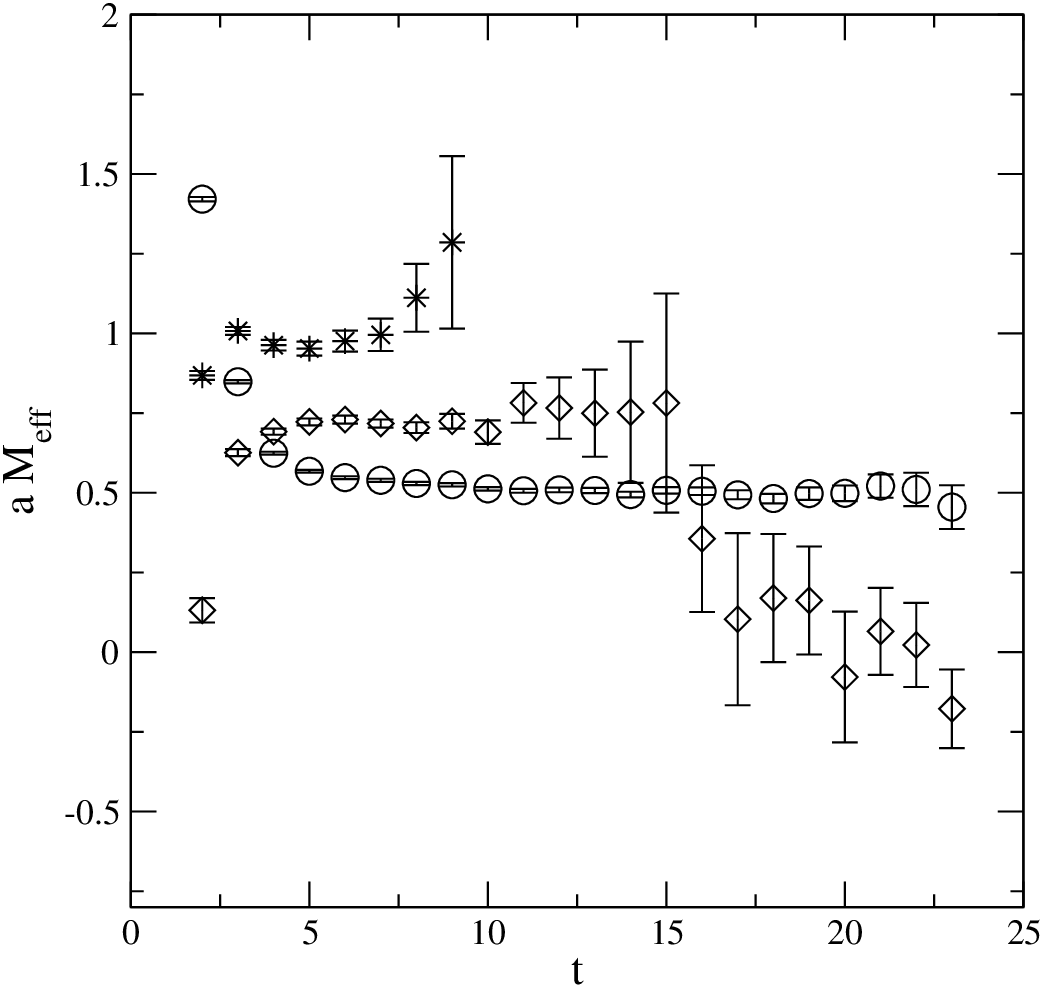}
\epsfxsize=140pt\hspace{0.5in}\epsfbox{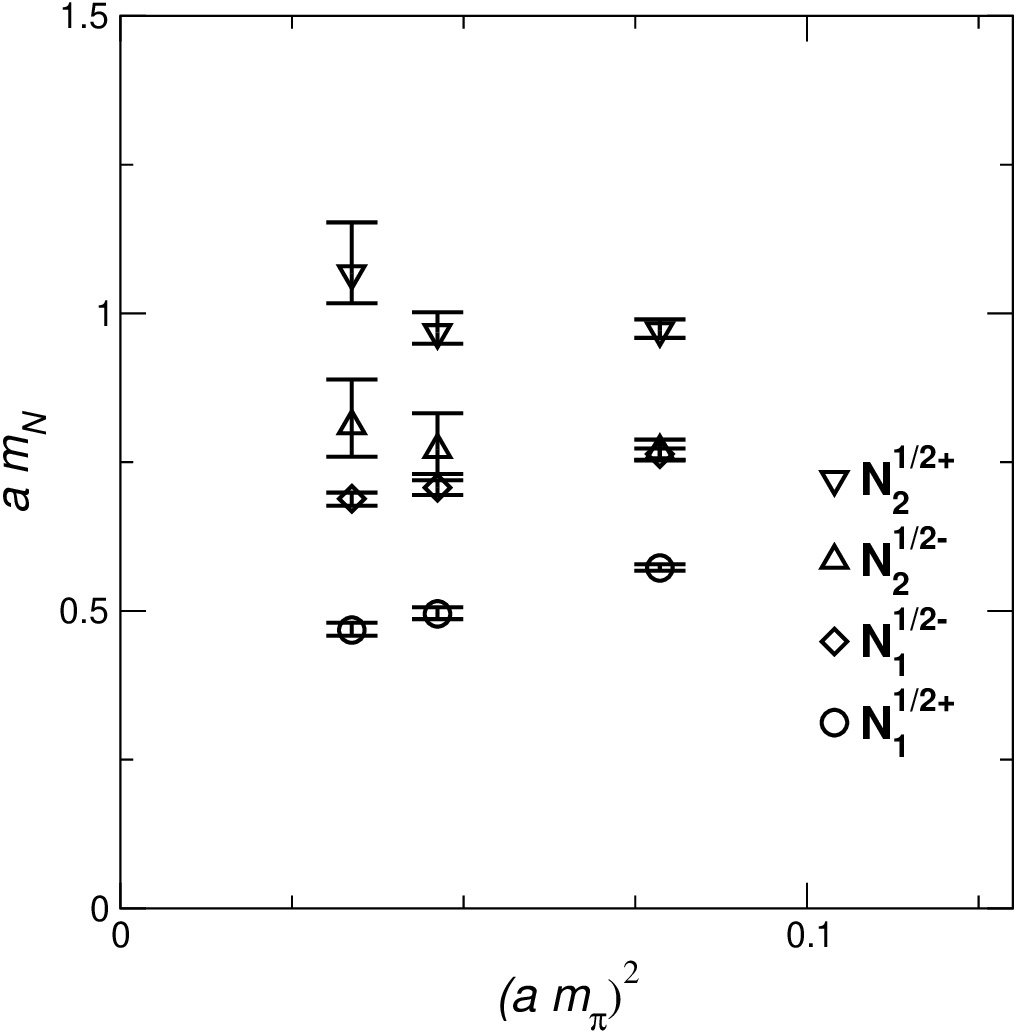} 
\caption{The left-hand plot shows the effective masses of the
positive-parity states using $N_1$ (circles) and $N_2$ (bursts), and
negative-parity using $N_1$ (diamonds).
The right-hand plot shows
the corresponding fitted masses at $(a/r_0)^2 \sim 0.02$, the middle
points in Figure~\protect\ref{fig:extrap}.}
\label{fig:roper}
\end{center}
\end{figure}

The nature of the Roper, the first nucleon excitation, has long been
debated. In Figure~\ref{fig:roper}, we show the effective masses of
the positive- and negative-parity states constructed from $N_1$, and
of the positive-parity state constructed using $N_2$ for a quark mass
around that of the strange; it is clear that the latter mass is
considerably higher than that of the negative-parity state, and
therefore much heavier than the Roper (1440).  The ordering of the
masses at each quark mass is also shown in the figure, revealing a
mass splitting between the radial excitation and the nucleon parity
partner comparable to that between the parity partner and the nucleon,
in accord with other lattice calculations.~\cite{lee98,lee00,sasaki99}

\section{Conclusions}
We have seen that the low-lying excited nucleon spectrum is accessible
to lattice calculation, and that lattice calculations are already
providing valuable insight, most notably through the lack of evidence
for the Roper resonance as a naive three-quark state.  Increasingly
energetic excitations are subject to increasing statistical noise, and thus
further precise calculations will require the full panoply of lattice
technology, such as the use of anisotropic lattices.~\cite{lee00,lhpc02}
Such lattice calculations will provide the vital theoretical
complement to the experimental programme at Jefferson Laboratory and
elsewhere.

\section*{Acknowledgements}
This work was supported in part by DOE contract DE-AC05-84ER40150
under which the Southeastern Universities Research Association (SURA)
operates the Thomas Jefferson National Accelerator Facility, by EPSRC
grant GR/K41663, and PPARC grants GR/L29927, GR/L56336 and
PPA/P/S/1998/00255.  MG acknowledges financial support from the DFG
(Schwerpunkt ``Elektromagnetische Sonden'').

\end{document}